\documentclass[sigconf,nonacm]{acmart}
\usepackage{enumitem}
\usepackage{hyperref}
\usepackage{multirow}
\usepackage{amsthm}
\usepackage{array}
\usepackage{subfigure}
\usepackage{graphicx}
\setlist[enumerate]{align=left}
\usepackage{algpseudocode}
\usepackage{algorithm}
\usepackage{amsthm} 

\usepackage{subcaption}

\usepackage[multiple]{footmisc}

\AtBeginDocument{%
  }
\setcopyright{acmlicensed}
\copyrightyear{2027}
\acmYear{2027}
\acmDOI{XXXXXXX.XXXXXXX}
\setcopyright{none}
\acmISBN{978-1-4503-XXXX-X/2018/06}

\begin{document}
\title{DuTri: Dual-Branch Modeling with Triadic Factorization for Multimodal Multi-Scenario Recommendation}

\author{Jintao Huang}
\affiliation{%
  \institution{College of Computer Science and Software Engineering, Shenzhen University}
  \city{Shenzhen}
  \country{China}
}
\email{jintao.huang35@gmail.com}

\author{Chaohua Yang}
\affiliation{%
  \institution{College of Computer Science and Software Engineering, Shenzhen University}
  \city{Shenzhen}
  \country{China}
}
\email{chaohua.ych@gmail.com}

\author{Kailiang Hao}
\affiliation{%
  \institution{College of Computer Science and Software Engineering, Shenzhen University}
  \city{Shenzhen}
  \country{China}
}
\email{kailianghao12@gmail.com}

\author{Dugang Liu}
\affiliation{%
  \institution{College of Computer Science and Software Engineering, Shenzhen University}
  \city{Shenzhen}
  \country{China}}  
\email{dugang.ldg@gmail.com}

\author{Xing Tang}
\affiliation{
  \institution{Shenzhen Technology University}
  \city{Shenzhen}
  \country{China}
}
\email{xing.tang@hotmail.com}

\author{Zhong Ming}
\affiliation{%
  \institution{College of Computer Science and Software Engineering, Shenzhen University}
  \city{Shenzhen}
  \country{China}
}
\email{mingz@szu.edu.cn}

\renewcommand{\shortauthors}{Jintao Huang, et al.}

\begin{abstract}
As recommendation scenarios become increasingly diverse and multimodal information grows more prevalent, effectively modeling cross-scenario user preferences and multimodal semantics has become a critical challenge in recommender systems. When incorporating multimodal information, existing multi-scenario recommendation (MSR) methods typically treat textual and visual representations as generic input features and directly feed them into their original parameter-sharing architectures. However, this extension may not fully accommodate the different modeling requirements of collaborative and multimodal information and may not be the best modeling paradigm. Our controlled experiments show that directly adopting these architectures is not the optimal modeling approach. To this end, we propose Dual-Branch Modeling with Triadic Factorization (DuTri), a framework for multimodal multi-scenario recommendation comprising a collaborative branch and a multimodal branch. Specifically, the collaborative branch learns structured collaborative preference representations through CANDECOMP/PARAFAC (CP) decomposition of user–scenario–item interactions and residual quantization, thereby capturing cross-scenario commonalities and scenario-specific differences. The multimodal branch leverages the learned user and scenario preferences to guide adaptive modality fusion; additionally, item-level contrastive learning aligns collaborative item preference representations with multimodal representations to facilitate effective collaboration between the two branches. Extensive experiments on two real-world datasets, AntM$^2$C and Amazon, demonstrate that DuTri outperforms existing multimodal and multi-scenario recommendation baselines in overall recommendation performance, validating the effectiveness of the proposed framework.
\end{abstract}





\maketitle

\section{Introduction}
\label{sec:intro}

Deep Recommender Systems (DRS) have been widely deployed on online platforms, where they alleviate information overload and improve user experience. As internet services evolve into increasingly diversified commercial ecosystems, industry has a growing demand for a unified recommender that can serve multiple scenarios with different user behaviors and data distributions~\cite{sheng2021one}. This trend has motivated research on multi-scenario recommendation (MSR), which aims to jointly model cross-scenario commonalities and scenario-specific characteristics within a unified framework~\cite{sheng2021one,xu2023musenet,zhang2022leaving,tang2023optmsm}. Meanwhile, multimodal information, such as item titles and images, provides rich semantic signals for modeling item characteristics and user preferences, and has demonstrated effectiveness in click-through rate (CTR) prediction~\cite{li2020adversarial,sheng2024enhancing,fan2025decoupled,cui2025diffusion}. By capturing content-level semantics beyond sparse ID features, such information can further facilitate knowledge transfer and improve generalization~\cite{hou2023learning,singh2024better}. Therefore, incorporating multimodal information into multi-scenario recommender systems is important for improving industrial recommendation performance.
\begin{figure}[htbp]
    \centering
    \includegraphics[width=1\linewidth]{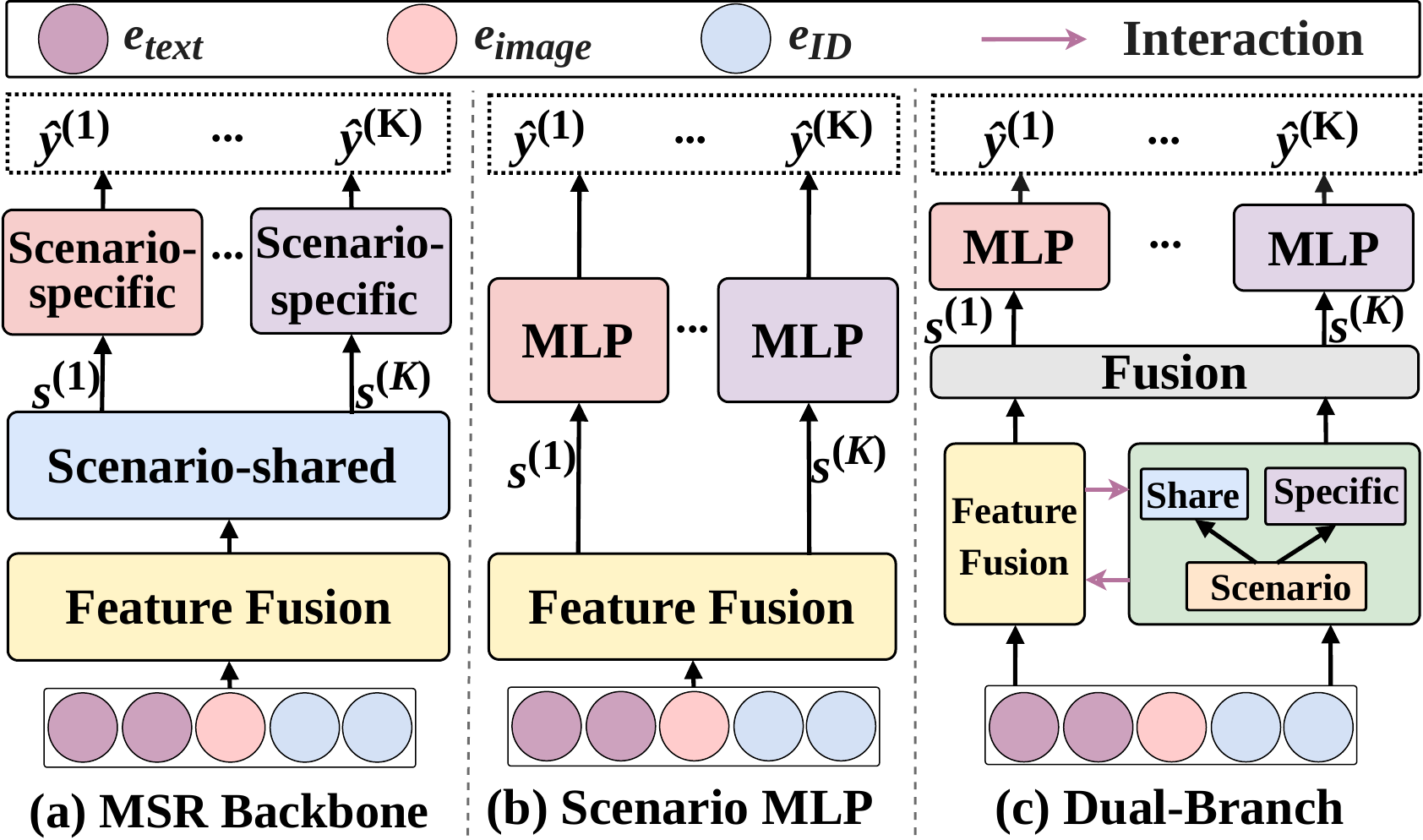}
    \Description[<Figure 1. Fully described in the text.>]{<A full description of Figure 1 can be found in the second paragraph of Section 1.>}
    \caption{Multimodal modeling strategies for multi-scenario recommendation: (a) MSR backbones with early fusion of multimodal features, (b) Scenario MLP with early fusion of multimodal features and scenario-specific prediction networks, and (c) a dual-branch modeling paradigm.}
    \label{fig:motivation}
\end{figure}

\begin{figure}[th]
    \centering
    \includegraphics[width=1\linewidth]{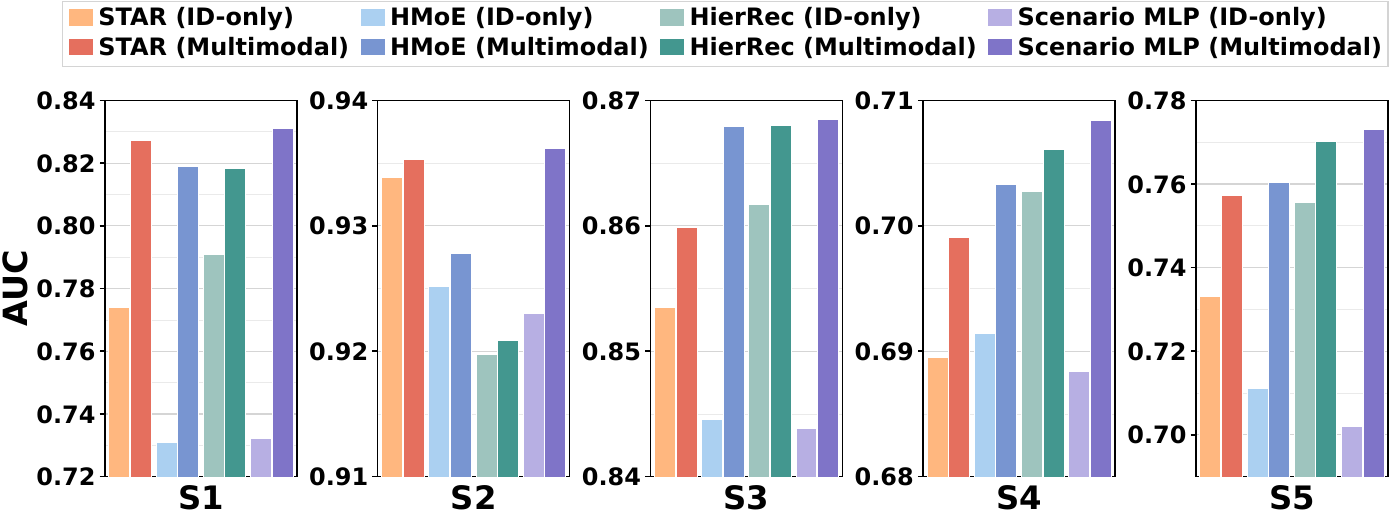}
    \Description[<Figure 1. Fully described in the text.>]{<A full description of Figure 2 can be found in the second paragraph of Section 1.>}
    \caption{Comparison of MSR backbones and Scenario MLP under ID-only and multimodal inputs on AntM$^2$C. Results report AUC for each scenario (S1--S5), averaged over five random seeds.}
    \label{fig:case}
    \vspace{-5pt}
\end{figure}

Integrating multimodal information into multi-scenario recommendation requires capturing cross-scenario shared preferences and scenario-specific variations while effectively leveraging multimodal semantics. When incorporating multimodal information, existing multi-scenario recommendation methods often treat it as generic input features and directly reuse their original parameter-sharing architectures to model cross-scenario commonalities and scenario-specific characteristics~\cite{zhao2025joint}. However, multimodal and collaborative information differ in their representation spaces and semantic content, and directly reusing the original parameter-sharing architectures may not fully accommodate their different modeling requirements, making this extension potentially suboptimal. To examine the effectiveness of this extension, we conduct controlled experiments comparing conventional MSR backbones with scenario-specific prediction networks under ID-only and multimodal settings. As illustrated in Fig.~\ref{fig:motivation}, we extend three representative MSR models (STAR~\cite{sheng2021one}, HMoE~\cite{li2020improving}, and HierRec~\cite{gao2024hierrec}) with multimodal features while preserving their original parameter-sharing structures (Fig.~\ref{fig:motivation}(a)). For comparison, we replace their MSR backbones with scenario-specific MLPs (Fig.~\ref{fig:motivation}(b)), while retaining shared ID embeddings and feature-fusion layers. Both strategies are evaluated under ID-only and multimodal input settings, with identical input features, fusion methods, and training and evaluation configurations within each setting. As shown in Fig.~\ref{fig:case}, the evaluated MSR models outperform Scenario MLP under ID-only inputs on AntM$^2$C, indicating that their cross-scenario sharing mechanisms offer advantages in this setting. However, after introducing multimodal information, Scenario MLP achieves higher mean AUC than all three MSR models across every scenario. A similar trend is observed on Amazon, as reported in Appendix~\ref{app:motivation_amazon}. These results indicate that the advantages of conventional cross-scenario sharing mechanisms under ID-only inputs do not necessarily carry over to multimodal settings, suggesting that directly reusing the original parameter-sharing architectures may not be the best modeling strategy. This motivates us to explore a dual-branch architecture in which the collaborative branch learns collaborative preferences that capture cross-scenario commonalities and scenario-specific variations, while the multimodal branch uses these preferences to guide adaptive multimodal fusion (Fig.~\ref{fig:motivation}(c)).

To this end, we propose Dual-Branch Modeling with Triadic Factorization (DuTri). This framework comprises interconnected collaborative and multimodal branches responsible for collaborative preference learning and multimodal fusion, respectively (Fig.~\ref{fig:motivation}(c)). Unlike approaches that directly reuse existing parameter-sharing architectures, DuTri learns collaborative preferences from user--scenario--item interactions while modeling multimodal information through a complementary multimodal branch. Specifically, the collaborative branch employs CANDECOMP/PARAFAC (CP) decomposition and residual quantization to learn structured collaborative preference representations. By sharing user and item factors across scenarios and using scenario factors to modulate their interactions, this branch captures cross-scenario commonalities and scenario-specific variations within collaborative preference learning. Meanwhile, the multimodal branch leverages the learned user and scenario preferences to guide adaptive fusion of item ID, textual, and visual features, allowing modality contributions to vary across users and scenarios. Furthermore, item-level contrastive learning aligns collaborative item preference representations with multimodal representations, facilitating information exchange between the two branches. The resulting representations are used for scenario-specific prediction. Our main contributions are as follows:

\begin{itemize}[leftmargin=*]
    \item We identify a limitation of directly reusing existing parameter-sharing architectures in multimodal multi-scenario recommendation: conventional cross-scenario sharing mechanisms that are effective under ID-only inputs may lose their advantages when applied to fused multimodal representations, motivating a reconsideration of how collaborative preference learning and multimodal fusion should be organized.

    \item We propose DuTri, an interconnected dual-branch framework that learns structured collaborative preference representations directly from user--scenario--item interactions through CP-based triadic factorization and residual quantization, while capturing cross-scenario commonalities and scenario-specific variations. The learned preferences further guide adaptive multimodal fusion and facilitate item-level cross-branch alignment.

    \item Experiments on AntM$^2$C and Amazon demonstrate that DuTri outperforms the evaluated multimodal and multi-scenario recommendation baselines, with ablation studies further validating the effectiveness of its key components.
\end{itemize}

\section{Related Work}
\label{sec:related}

We review two lines of research most relevant to our work: multi-scenario recommendation and multimodal recommender systems.

\noindent\textbf{Multi-scenario Recommendation.}
Multi-scenario recommendation (MSR) jointly models multiple business scenarios that differ in user groups and data distributions~\cite{li2020improving,sheng2021one,li2022adaptdhm,shen2021sar,zhang2022leaving}. Early studies mainly focus on balancing scenario commonality and specificity. For example, HMoE~\cite{li2020improving} models task relationships implicitly in the feature space and explicitly in the label space, while STAR~\cite{sheng2021one} pairs a global center module with scenario-specific branches. More recent work explores finer-grained scenario relationships. M2M~\cite{zhang2022leaving} models inter-scenario dependencies with meta-attention, AdaptDHM~\cite{li2022adaptdhm} dynamically assigns samples to latent domains, DFFM~\cite{DFFM} models scenario-aware feature interactions, and HierRec~\cite{gao2024hierrec} captures hierarchical correlations among scenarios. Other studies investigate scenario commonality from the data and feature perspectives, such as adaptive selection of shared instances in SSIM~\cite{ssim} and instance-level embedding optimization in MultiEmb~\cite{yang2025multi}. Most of these methods are designed around ID-based features and do not explicitly address multimodal fusion. Related context-aware recommendation methods, such as iTALS~\cite{hidasi2012fast}, use tensor factorization to model user--item--context interactions. AMS~\cite{hao2026ams} generates sample-level modality activation masks conditioned on scenario and modality information, and decouples its modality fusion module from the MSR backbone. However, iTALS does not address multimodal fusion, while AMS focuses on modality scheduling without explicitly aligning modality representations with separately modeled collaborative preferences. DuTri organizes cross-scenario sharing through shared user and item preference factors and scenario factors that modulate their interactions, and uses this preference structure to connect separate collaborative and multimodal branches.

\noindent\textbf{Multimodal Recommendation.}
Multimodal recommendation methods draw on textual, visual, and other heterogeneous signals to improve user-interest modeling. These methods largely center on cross-modal interaction and alignment, efficient modality fusion, and the transfer of semantic knowledge~\cite{li2020adversarial,sheng2024enhancing,qiu2024ease,fan2025decoupled,wang2025progressive,liu2023large}. MARN~\cite{li2020adversarial} separates modality-shared representations from modality-specific ones, whereas GMMF~\cite{xiao2022abstract} derives user-specific fusion weights. MAKE~\cite{sheng2024enhancing} and EASE~\cite{qiu2024ease} enhance representation generalization through semantic-aware pre-training and LLM-guided adapters, respectively. DMF~\cite{fan2025decoupled} models fine-grained interactions between ID-based collaborative representations and multimodal representations, while Diff-MSIN~\cite{cui2025diffusion} explores cross-modal synergy and interactions. SOLID~\cite{lv2024semantic} and PSRQ~\cite{wang2025progressive} adopt semantic codebooks or residual quantization to learn multimodal representations. However, these studies primarily focus on multimodal representation learning and fusion. DuTri explicitly incorporates cross-scenario collaborative preferences into multimodal modeling: user and scenario factors jointly condition the fusion of ID, textual, and visual representations, while item factors support alignment between collaborative preferences and modality representations.
\section{Problem Formulation}

In this section, we formalize the multimodal multi-scenario recommendation task and introduce the notation used throughout the paper. Let $\mathcal{S}=\{s_1,\ldots,s_K\}$ denote the set of $K$ recommendation scenarios, and $\mathcal{M}=\{\mathrm{ID},T,V\}$ denote the set of item modalities, including item identifiers, textual information (e.g., titles and descriptions), and visual content (e.g., main images). User identifiers are modeled separately from item modalities. Let $\mathcal{X}$ denote the feature space shared across scenarios, and $\mathcal{Y}=\{0,1\}$ denote the binary label space, where $1$ and $0$ indicate positive and negative feedback, respectively. Each interaction involves a user $u$, represented by the identifier $x_u$, and an item $i$, represented by $x_i=\{x_i^{\mathrm{ID}},x_i^T,x_i^V\}$. Here, $x_i^{\mathrm{ID}}$ denotes the item identifier, while $x_i^T\in\mathbb{R}^{d_T}$ and $x_i^V\in\mathbb{R}^{d_V}$ denote textual and visual embeddings extracted offline using pretrained encoders. For each scenario $s_k$, the training set is defined as $\mathcal{D}_k=\{(x_j^k,y_j^k,s_k)\}_{j=1}^{N_k}$, where $N_k$ is the number of instances in scenario $s_k$, $x_j^k\in\mathcal{X}$ comprises the user identifier $x_{u_j^k}$ and item modalities $x_{i_j^k}$ of the $j$-th instance, and $y_j^k\in\mathcal{Y}$ is the corresponding feedback label. The complete training set is $\mathcal{D}=\bigcup_{k=1}^{K}\mathcal{D}_k$, where the subsets $\mathcal{D}_k$ are pairwise disjoint, with each instance belonging to exactly one scenario.

Given these observations, the goal of multimodal multi-scenario recommendation is to learn a unified prediction function that captures cross-scenario shared and scenario-specific user--item preferences while effectively leveraging collaborative and multimodal information. For the $j$-th instance in scenario $s_k$, the predicted probability of positive feedback is given by
\begin{equation}
    \hat{y}_j^k=f(x_j^k,s_k\mid\theta,\{\theta_{s_k}\}),
\label{eq:general_prediction}
\end{equation}
where $\theta$ and $\theta_{s_k}$ denote the scenario-shared and scenario-specific model parameters, respectively. The model is trained by minimizing the average binary cross-entropy loss over all instances:
\begin{equation}
    \mathcal{L}_{\mathrm{Rec}}=\frac{1}{N}\sum_{k=1}^{K}\sum_{j=1}^{N_k}\ell\left(y_j^k,\hat{y}_j^k\right),\qquad N=\sum_{k=1}^{K}N_k,
\label{eq:rec_loss}
\end{equation}
where $\ell(\cdot)$ denotes the binary cross-entropy loss function.
\section{The Proposed Framework}
\label{sec:method}
In this section, we present the overall framework of DuTri. We first describe the overview of the framework, then detail the collaborative and multimodal branches, and finally the joint optimization.
\subsection{Framework Overview}
\label{sec:framework_overview}
\begin{figure*}[t]
    \centering
    \includegraphics[width=1.0\textwidth]{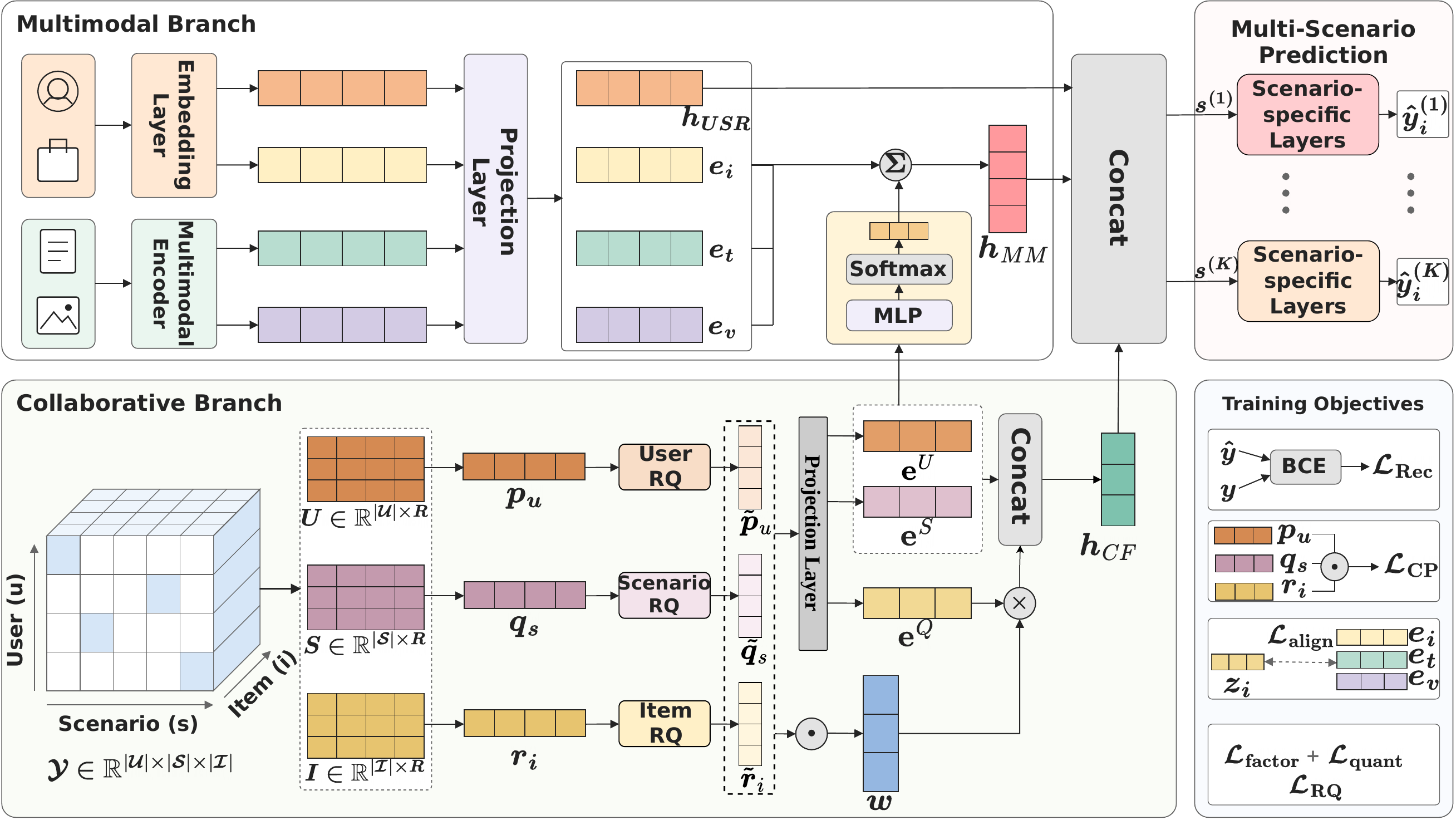}
    \Description[<Figure 3. Fully described in the text.>]{<A full description of Figure 3 can be found in Section 4.1.>}
    \caption{Overview of DuTri, a dedicated dual-branch framework for multimodal multi-scenario recommendation. Triadic preferences guide item-level alignment and user--scenario-conditioned fusion. Collaborative and multimodal representations are combined with the user ID representation for scenario-specific prediction.}
    \label{fig:framework}
    \vspace{-10pt}
\end{figure*}
As illustrated in Fig.~\ref{fig:framework}, DuTri consists of a collaborative branch and a multimodal branch coupled through preference representations. The collaborative branch factorizes user--scenario--item interactions with CP factorization, fitted to observed labels, and quantizes the resulting factors into preference representations. These representations feed the collaborative representation and condition the multimodal branch. An item-level contrastive objective aligns item preferences with modality representations during training. 
The outputs of both branches are concatenated with the user representation and passed to the prediction network corresponding to the current scenario.
\subsubsection{Modality Feature Embedding}
\label{sec:modality_initialization}
To better represent the identifier features, we map the user and item identifiers to dense embeddings using an embedding table shared across scenarios:
\begin{equation}
    \mathbf{r}_u^{\mathrm{ID}}=\mathbf{E}\times\mathbf{x}_u,\qquad \mathbf{r}_i^{\mathrm{ID}}=\mathbf{E}\times\mathbf{x}_i^{\mathrm{ID}},
\label{eq:id_embedding}
\end{equation}
where $\mathbf{E}\in\mathbb{R}^{d_{\mathrm{ID}}\times(|\mathcal{U}|+|\mathcal{I}|)}$ is the trainable embedding table belonging to $\theta$, and $\mathbf{x}_u$ and $\mathbf{x}_i^{\mathrm{ID}}$ are one-hot vectors over the joint index set of users and items. Each identifier field is looked up separately, and the two embeddings are kept apart rather than summed. For textual and visual inputs, the raw contents of item $i$ are encoded by pretrained encoders:
\begin{equation}
    \mathbf{r}_i^{T}=\operatorname{TextEncoder}(\mathbf{C}_i^{T}),\qquad \mathbf{r}_i^{V}=\operatorname{ImageEncoder}(\mathbf{C}_i^{V}),
\label{eq:modality_encoders}
\end{equation}
where $\mathbf{C}_i^{T}$ and $\mathbf{C}_i^{V}$ denote the raw title-and-description text and the raw images of item $i$. The encoders are frozen and applied offline, so the resulting embeddings $\mathbf{r}_i^{T}\in\mathbb{R}^{d_T}$ and $\mathbf{r}_i^{V}\in\mathbb{R}^{d_V}$ are fixed during training. Details of the encoders are given in Appendix~\ref{app:datasets}. Each item-side modality $m\in\mathcal{M}$ is projected into a common $d$-dimensional space with its own parameters:
\begin{equation}
    \mathbf{e}^{m}=\mathbf{W}_{m}\mathbf{r}_i^{m}+\mathbf{b}_{m},
\label{eq:modality_projection}
\end{equation}
where $\mathbf{W}_m$ and $\mathbf{b}_m$ are the weight matrix and bias specific to modality $m$. The user identifier is projected separately:
\begin{equation}
    \mathbf{h}_{\mathrm{USR}}=\mathbf{W}_{\mathrm{USR}}\mathbf{r}_{u}^{\mathrm{ID}}+\mathbf{b}_{\mathrm{USR}},
\label{eq:user_projection}
\end{equation}
where $\mathbf{W}_{USR}$ and $\mathbf{b}_{USR}$ are the weight matrix and bias. The projection parameters are trainable and shared across scenarios. The user representation is kept out of the item-side fusion and enters only the final prediction.
\subsubsection{Collaborative Branch Overview}
The collaborative branch models user--scenario--item interactions with CP factorization. Its user, scenario, and item factors are quantized into preference representations $\widetilde{\mathbf{p}}$, $\widetilde{\mathbf{q}}$, and $\widetilde{\mathbf{r}}$, which are combined into a collaborative representation:
\begin{equation}
    \mathbf{h}_{\mathrm{CF}}=\operatorname{CF}\left(\widetilde{\mathbf{p}},\widetilde{\mathbf{q}},\widetilde{\mathbf{r}}\right),
\label{eq:overview_collaborative}
\end{equation}
where $\operatorname{CF}$ denotes the triadic modulation and feed-forward transformation detailed in Sec.~\ref{sec:quantized_tri_modeling}. The user and scenario preferences also condition the multimodal branch through the gate.
\subsubsection{Multimodal Branch Overview}
The multimodal branch fuses the item-side representations of $\mathcal{M}$, with modality weights determined by the projected user and scenario preferences:
\begin{equation}
    \mathbf{h}_{\mathrm{MM}}=\operatorname{MM}\left(\mathbf{e}^{\mathrm{ID}},\mathbf{e}^{T},\mathbf{e}^{V}\mid\mathbf{e}^{U},\mathbf{e}^{S}\right),
\label{eq:multimodal_representation}
\end{equation}
where $\operatorname{MM}$ denotes the gating and fusion detailed in Sec.~\ref{sec:multimodal_branch}, and $\mathbf{e}^{U}$ and $\mathbf{e}^{S}$ are the user and scenario projections defined in Eq.~\eqref{eq:preference_projection}. During training, an item-level contrastive objective aligns the collaborative item preference with each modality representation.
\subsubsection{Prediction}
For an instance from scenario $s_k$, the user, collaborative, and multimodal representations are concatenated and passed to the scenario-specific MLP. The prediction heads share the same architecture but have separate parameters. Scenario information influences representation construction through the scenario factor and determines which prediction head is used. Training combines $\mathcal{L}_{\mathrm{Rec}}$ with the auxiliary terms detailed in Sec.~\ref{sec:joint_optimization}.
\subsection{Collaborative Branch}
\label{sec:quantized_tri_modeling}
\subsubsection{Triadic Factorization of Preference Factors}
To capture collaborative preferences across scenarios, we factorize user--scenario--item interactions with CP factorization. For user $u$, scenario $s_k$, and item $i$, we introduce preference factors $\mathbf{p}_u,\mathbf{q}_k,\mathbf{r}_i\in\mathbb{R}^{d_c}$, where $d_c$ is the factor dimension. These factors are parameterized separately from the ID embedding $\mathbf{E}$. The user and item factors are shared across scenarios and belong to $\theta$, whereas the scenario factor $\mathbf{q}_k$ belongs to $\theta_{s_k}$. The CP interaction score is
\begin{equation}
    a_{u,k,i}=\sum_{\rho=1}^{d_c}p_{u,\rho}\,q_{k,\rho}\,r_{i,\rho}.
\label{eq:cp_interaction_score}
\end{equation}
We treat $a_{u,k,i}$ as an auxiliary logit, map it to a probability with the Sigmoid function, and minimize the binary cross-entropy over a mini-batch $\mathcal{B}$ of interaction instances, which contain both positive and negative labels:
\begin{equation}
    \mathcal{L}_{\mathrm{CP}}=\frac{1}{B}\sum_{(u,s_k,i,y)\in\mathcal{B}}\ell\left(y,\operatorname{Sigmoid}(a_{u,k,i})\right),\qquad B=|\mathcal{B}|,
\label{eq:cp_supervision_loss}
\end{equation}
where $\ell(\cdot)$ is the binary cross-entropy loss. Since $\mathcal{L}_{\mathrm{CP}}$ acts on the continuous factors before quantization, the quantized preferences are not directly supervised by it.

To encourage structured and regularized preference representations, we further apply residual quantization (RQ) to the learned factors. By representing continuous preferences through discrete codewords, RQ encourages the reuse of latent preference patterns and provides regularized representations for subsequent modeling. The learned preference factors are quantized and reconstructed
using three entity-specific residual quantizers:
\begin{equation}
\widetilde{\mathbf p}_u=\mathcal Q_U(\mathbf p_u),\quad
\widetilde{\mathbf q}_k=\mathcal Q_S(\mathbf q_k),\quad
\widetilde{\mathbf r}_i=\mathcal Q_I(\mathbf r_i).
\label{eq:triadic_factor_discretization}
\end{equation}
Each quantizer has its own encoder, residual codebooks,
and linear decoder, with $L_U$, $L_S$, and $L_I$
quantization levels, respectively.
The selected codewords are summed and decoded into
continuous preference representations for subsequent modeling. These quantizers are independently parameterized and shared across scenarios. To enable end-to-end optimization despite discrete codeword selection, we adopt the straight-through estimator, allowing gradients to propagate through the quantization process. The resulting reconstructed preferences are used in subsequent modeling, with the quantization objective \(\mathcal{L}_{\mathrm{RQ}}\) detailed in Appendix~\ref{app:quantization_details}.

\subsubsection{Preference-Modulated Item Representation} 
The reconstructed preference factors are projected into a common $d$-dimensional space:
\begin{equation}
\mathbf e^U=\mathbf W_U\widetilde{\mathbf p}_u+\mathbf b_U,\quad
\mathbf e^S=\mathbf W_S\widetilde{\mathbf q}_k+\mathbf b_S,\quad
\mathbf e^Q=\mathbf W_Q\widetilde{\mathbf r}_i+\mathbf b_Q.
\label{eq:preference_projection}
\end{equation}
Each projection uses separate trainable parameters.
The user preference projection $\mathbf e^U$ is derived from the CP user factor and is distinct from the ID-based user representation $\mathbf h_{\mathrm{USR}}$ defined in Sec.~\ref{sec:modality_initialization}. The three reconstructed preferences interact multiplicatively to produce a feature-wise modulation vector, which rescales the projected item preference:
\begin{equation}
    \mathbf{c}=\operatorname{LN}\left(
    \mathbf{e}^{Q}\odot
    f_{\mathrm{tri}}\left(
    \widetilde{\mathbf{p}}_u\odot
    \widetilde{\mathbf{q}}_k\odot
    \widetilde{\mathbf{r}}_i
    \right)\right),
\label{eq:triadic_modulation}
\end{equation}
where $\odot$ denotes element-wise multiplication, $f_{\mathrm{tri}}:\mathbb{R}^{d_c}\rightarrow\mathbb{R}^{d}$ consists of a linear layer, layer normalization, and GELU, and $\mathbf{c}\in\mathbb{R}^{d}$ depends on the user, scenario, and item preferences jointly. The collaborative representation concatenates the user and scenario projections with $\mathbf{c}$:
\begin{equation}
    \mathbf{h}_{\mathrm{CF}}=\operatorname{FFN}_{\mathrm{CF}}\left([\mathbf{e}^{U}\Vert\mathbf{e}^{S}\Vert\mathbf{c}]\right),
\label{eq:collaborative_representation}
\end{equation}
where $\operatorname{FFN}_{\mathrm{CF}}$ consists of a linear layer, layer normalization, and GELU. The projections $\mathbf{e}^{U}$ and $\mathbf{e}^{S}$ also serve as inputs to the gating network in Sec.~\ref{sec:multimodal_branch}, and $\mathbf{e}^{Q}$ is aligned with the modality representations in Sec.~\ref{sec:joint_optimization}.
\subsection{Multimodal Branch}
\label{sec:multimodal_branch}
The multimodal branch fuses the item-side modalities. Since the relative importance of these modalities may vary across users and scenarios, the branch determines their weights from the projected user and scenario preferences.
\subsubsection{Conditional Weight Generation}
The projections $\mathbf{e}^{U}$ and $\mathbf{e}^{S}$ from Eq.~\eqref{eq:preference_projection} are concatenated and passed to a gating network:
\begin{equation}
    \boldsymbol{\alpha}=\operatorname{Softmax}\left(g_{\mathrm{gate}}\left([\mathbf{e}^{U}\Vert\mathbf{e}^{S}]\right)\right),
\label{eq:conditional_modality_weights}
\end{equation}
where $g_{\mathrm{gate}}:\mathbb{R}^{2d}\rightarrow\mathbb{R}^{|\mathcal{M}|}$ consists of a linear layer, layer normalization, GELU, and an output linear layer. The weights satisfy $\sum_{m\in\mathcal{M}}\alpha^m=1$. The gate is shared across scenarios and takes only the user and scenario preferences as input.
\subsubsection{Conditional Multimodal Fusion}
To reduce the influence of scale differences across modalities, each item-side representation is normalized before weighted aggregation:
\begin{equation}
    \mathbf{h}_{\mathrm{MM}}=\operatorname{LN}_{O}\left(\sum_{m\in\mathcal{M}}\alpha^{m}\operatorname{LN}_{m}(\mathbf{e}^{m})\right),
\label{eq:conditional_multimodal_fusion}
\end{equation}
where $\operatorname{LN}_{m}$ is the modality-specific layer normalization, and $\operatorname{LN}_{O}$ normalizes the fused representation. The resulting $\mathbf{h}_{\mathrm{MM}}\in\mathbb{R}^{d}$ enters the final prediction alongside $\mathbf{h}_{\mathrm{CF}}$. 
Within the representation-learning modules, all trainable parameters except the scenario factors $\mathbf{q}_k$ are shared across scenarios.
\subsection{Optimization}
\label{sec:joint_optimization}
To encourage consistency between collaborative preferences and multimodal representations, we introduce an item-level contrastive alignment objective. For each instance $b$ in a mini-batch $\mathcal{B}$, the collaborative item preference $\mathbf{e}_b^Q$ and each modality representation $\mathbf{e}_b^m$ are projected into a shared space and $\ell_2$-normalized:
\begin{equation}
    \mathbf{z}_b^Q=\operatorname{Normalize}(g^Q(\mathbf{e}_b^Q)),
    \quad
    \mathbf{z}_b^m=\operatorname{Normalize}(g^m(\mathbf{e}_b^m)),
\label{eq:alignment_embedding}
\end{equation} 
where $g^Q$ is the shared collaborative projection head, $g^m$ is the modality-specific projection head, and $\operatorname{Normalize}(\cdot)$ denotes $\ell_2$ normalization. The modality representations are taken before the normalization used for fusion. For each anchor $b$, instances associated with the same item form the positive set
$\mathcal{P}(b)=\{a\in\mathcal{B}\mid i_a=i_b\}$. The contrastive alignment loss for modality $m\in\mathcal{M}$ is
\begin{equation}
    \mathcal{L}_{\mathrm{align}}^m
    =-\frac{1}{|\mathcal{B}|}
    \sum_{b\in\mathcal{B}}
    \log
    \frac{
        \sum_{a\in\mathcal{P}(b)}
        \exp((\mathbf{z}_b^Q)^\top\mathbf{z}_a^m/\tau)
    }{
        \sum_{c\in\mathcal{B}}
        \exp((\mathbf{z}_b^Q)^\top\mathbf{z}_c^m/\tau)
    },
\label{eq:alignment_loss}
\end{equation}
where $\tau$ is the temperature parameter. All instances serve as candidates, including the anchor itself, allowing repeated items to contribute multiple positive pairs. The overall training objective combines recommendation, CP supervision, residual quantization, and contrastive alignment:
\begin{equation}
    \mathcal{L}
    =\mathcal{L}_{\mathrm{Rec}}
    +\lambda_{\mathrm{CP}}\mathcal{L}_{\mathrm{CP}}
    +\mathcal{L}_{\mathrm{RQ}}
    +\lambda_{\mathrm{cl}}
    \sum_{m\in\mathcal{M}}\mathcal{L}_{\mathrm{align}}^m,
\label{eq:full_training_objective}
\end{equation}
where $\lambda_{\mathrm{CP}}$ and $\lambda_{\mathrm{cl}}$ control the contributions of CP supervision and contrastive alignment, respectively. The quantization loss $\mathcal{L}_{\mathrm{RQ}}$ is detailed in Appendix~\ref{app:quantization_details}. During training, alignment gradients propagate through both branches, with the straight-through estimator enabling updates to the continuous preference factors. The codebooks are optimized through the quantization losses, while the pretrained textual and visual features remain frozen. At inference, the projection heads and auxiliary objectives are discarded.

\section{Experiments}
\label{sec:experiments}
In this section, we evaluate DuTri on two real-world multimodal multi-scenario datasets to answer the following research questions:
\begin{itemize}[leftmargin=*]
    \item \textbf{RQ1}: How does DuTri compare with existing methods in recommendation performance?
    \item \textbf{RQ2}: How does each key component contribute to the performance of DuTri?
    \item \textbf{RQ3}: How do different strategies for cross-scenario sharing and multimodal integration affect recommendation performance?
    \item \textbf{RQ4}: How does DuTri perform in terms of computational efficiency and hyperparameter sensitivity?
\end{itemize}

\subsection{Experimental Setup}
\label{sec:experiments:setup}

\subsubsection{Datasets}
Following the benchmark setup of AMS~\cite{hao2026ams}, we evaluate DuTri on two real-world datasets: AntM$^2$C~\cite{huan2024exploring} and Amazon~\cite{ni2019justifying}. AntM$^2$C contains click logs and multimodal item information from five industrial recommendation scenarios on Alipay. For Amazon, we construct three category-based scenarios using Sports and Outdoors, Movies and TV, and Books. We retain users with interactions in all three categories and convert ratings into binary feedback. AntM$^2$C uses a global chronological split, while Amazon is split chronologically within each user--scenario pair. The Amazon protocol enforces temporal ordering within each pair and does not impose a common temporal cutoff across scenarios. Dataset statistics, preprocessing procedures, and feature extraction settings are provided in Appendix~\ref{app:datasets}.

\subsubsection{Metrics}
To assess recommendation quality, we report the area under the ROC curve (AUC) for each scenario~\cite{lin2022adafs,wang2022autofield}. Overall performance is summarized using sample-weighted AUC (WAUC), with weights determined by the number of test instances in each scenario. Small improvements in AUC can be practically meaningful in large-scale CTR prediction~\cite{guo2017deepfm}.

\subsubsection{Baseline Methods}
We compare DuTri with representative baselines from two groups: (1) multimodal recommendation and fusion methods, including Diff-MSIN~\cite{cui2025diffusion}, LMF~\cite{liu2018efficient}, PSRQ~\cite{wang2025progressive}, SimCEN~\cite{li2024simcen}, and DMF~\cite{fan2025decoupled}; and (2) multi-scenario and multi-task CTR methods, including STAR~\cite{sheng2021one}, HMoE~\cite{li2020improving}, MMoE~\cite{ma2018modeling}, PLE~\cite{tang2020progressive}, AdaptDHM~\cite{li2022adaptdhm}, DFFM~\cite{DFFM}, HierRec~\cite{gao2024hierrec}, and SSIM~\cite{ssim}. The multimodal baselines are trained on interactions pooled across scenarios and evaluated within each scenario. Methods in the second group receive concatenated item ID, textual, and visual representations through early fusion, while retaining their original sharing and interaction mechanisms.

We additionally evaluate AMS~\cite{hao2026ams}, which explicitly addresses modality scheduling in multi-scenario CTR prediction. We also consider Scenario MLP as a structural reference: it retains shared ID embeddings and a shared multimodal fusion module while using a separate prediction network for each scenario. This reference supports the comparison of sharing strategies after multimodal fusion. Details of baseline adaptations and implementation sources are provided in Appendix~\ref{app:baselines}.

\subsubsection{Implementation Details}
\label{sec:implementation_details}
All methods use the same data splits, available user and item ID information, pre-extracted textual and visual features, and evaluation protocol. Hyperparameters are tuned independently for each method using validation WAUC, and the checkpoint with the highest validation WAUC is evaluated on the test set. Recommendation results are averaged over five random seeds. Table~\ref{tab:main_result} reports the mean and standard deviation across these runs; other recommendation results report means for brevity. Detailed configurations and hyperparameter search ranges are provided in Appendix~\ref{app:implementation}.

\subsection{RQ1: Performance Comparison}
\label{sec:experiments:rq1}
\begin{table*}[t]
\centering
\caption{Results on both datasets, reported as mean $\pm$ standard deviation over five random seeds. The best mean in each column is boldfaced and the second-best underlined. Ties at displayed precision are marked equally. $^{*}$ indicates a statistically significant improvement over the baseline with the highest mean in the column under a two-sided paired $t$-test ($p<0.05$).}
\label{tab:main_result}
\vspace{-5px}
\setlength{\tabcolsep}{1pt}
\renewcommand{\arraystretch}{1.0}
\resizebox{\textwidth}{!}{%
\begin{tabular}{@{}c*{10}{c}@{}}
\toprule
\multirow{3}{*}{Method} & \multicolumn{6}{c}{AntM$^2$C} & \multicolumn{4}{c}{Amazon} \\
\cmidrule(lr){2-7}\cmidrule(lr){8-11}
& \multicolumn{5}{c}{AUC$\uparrow$} & \multirow{2}{*}{WAUC$\uparrow$} & \multicolumn{3}{c}{AUC$\uparrow$} & \multirow{2}{*}{WAUC$\uparrow$} \\
& S1 & S2 & S3 & S4 & S5 & & S1 & S2 & S3 & \\
\midrule

Diff-MSIN & .8448$\pm$.0051 & .9087$\pm$.0007 & .8310$\pm$.0005 & .6016$\pm$.0043 & .7384$\pm$.0039 & .8189$\pm$.0009 & .7183$\pm$.0011 & .6768$\pm$.0024 & .7760$\pm$.0013 & .7266$\pm$.0011 \\
SimCEN & .8014$\pm$.0066 & .9275$\pm$.0019 & .8640$\pm$.0008 & .6902$\pm$.0029 & .7584$\pm$.0073 & .8507$\pm$.0014 & .7243$\pm$.0020 & .6505$\pm$.0094 & .7580$\pm$.0077 & .7221$\pm$.0029 \\
PSRQ & .8031$\pm$.0051 & .9261$\pm$.0006 & .8556$\pm$.0039 & .6913$\pm$.0004 & .7350$\pm$.0029 & .8443$\pm$.0024 & .6945$\pm$.0070 & .6567$\pm$.0040 & .7660$\pm$.0015 & .7067$\pm$.0052 \\
LMF & .7960$\pm$.0059 & .9321$\pm$.0013 & .8682$\pm$.0013 & .7073$\pm$.0005 & .7584$\pm$.0070 & .8556$\pm$.0011 & .7364$\pm$.0003 & \underline{.6797}$\pm$.0001 & .7794$\pm$.0002 & .7389$\pm$.0002 \\
DMF & .7869$\pm$.0022 & .9269$\pm$.0007 & .8649$\pm$.0026 & .7023$\pm$.0011 & .7332$\pm$.0010 & .8504$\pm$.0017 & .7336$\pm$.0000 & .6755$\pm$.0002 & .7781$\pm$.0001 & .7363$\pm$.0000 \\
\hline
MMoE & .8150$\pm$.0067 & .9222$\pm$.0014 & .8585$\pm$.0016 & .6985$\pm$.0034 & .7679$\pm$.0048 & .8483$\pm$.0011 & .7350$\pm$.0001 & .6765$\pm$.0003 & .7787$\pm$.0002 & .7375$\pm$.0001 \\
PLE & .8241$\pm$.0151 & .9221$\pm$.0009 & .8615$\pm$.0013 & .7010$\pm$.0011 & .7702$\pm$.0041 & .8508$\pm$.0009 & .7349$\pm$.0001 & .6759$\pm$.0003 & .7787$\pm$.0001 & .7373$\pm$.0000 \\
HMoE & .8190$\pm$.0170 & .9278$\pm$.0025 & .8679$\pm$.0022 & .7033$\pm$.0026 & .7604$\pm$.0078 & .8553$\pm$.0013 & .7305$\pm$.0012 & .6573$\pm$.0128 & .7666$\pm$.0090 & .7289$\pm$.0042 \\
STAR & .8274$\pm$.0077 & .9353$\pm$.0008 & .8599$\pm$.0019 & .6991$\pm$.0023 & .7572$\pm$.0038 & .8517$\pm$.0016 & .7294$\pm$.0024 & .6681$\pm$.0033 & .7685$\pm$.0023 & .7303$\pm$.0020 \\
AdaptDHM & .7684$\pm$.0010 & .9129$\pm$.0001 & .8517$\pm$.0001 & .6893$\pm$.0001 & .7049$\pm$.0004 & .8360$\pm$.0001 & .7310$\pm$.0002 & .6718$\pm$.0001 & .7739$\pm$.0002 & .7331$\pm$.0002 \\
DFFM & \underline{.8595}$\pm$.0007 & .9222$\pm$.0007 & .8696$\pm$.0014 & .7072$\pm$.0001 & .7708$\pm$.0018 & .8579$\pm$.0008 & .7241$\pm$.0008 & .6650$\pm$.0007 & .7672$\pm$.0004 & .7263$\pm$.0006 \\
HierRec & .8182$\pm$.0077 & .9209$\pm$.0023 & .8680$\pm$.0034 & .7061$\pm$.0009 & .7701$\pm$.0017 & .8548$\pm$.0025 & .7344$\pm$.0007 & .6732$\pm$.0010 & .7773$\pm$.0004 & .7363$\pm$.0004 \\
SSIM & .8454$\pm$.0215 & \underline{.9364}$\pm$.0045 & .8636$\pm$.0013 & .7044$\pm$.0033 & .7716$\pm$.0094 & .8562$\pm$.0026 & .7361$\pm$.0001 & .6783$\pm$.0003 & .7794$\pm$.0001 & .7385$\pm$.0001 \\
Scenario MLP & .8311$\pm$.0042 & .9362$\pm$.0022 & .8685$\pm$.0012 & \underline{.7084}$\pm$.0013 & .7732$\pm$.0024 & .8590$\pm$.0014 & .7374$\pm$.0009 & .6794$\pm$.0005 & .7815$\pm$.0007 & \underline{.7400}$\pm$.0007 \\
\hline
AMS & .8555$\pm$.0046 & .9348$\pm$.0067 & \underline{.8697}$\pm$.0025 & .7010$\pm$.0038 & \underline{.7739}$\pm$.0101 & \underline{.8598}$\pm$.0007 & \underline{.7385}$\pm$.0002 & .6718$\pm$.0025 & \underline{.7832}$\pm$.0001 & \underline{.7400}$\pm$.0003 \\
\midrule
DuTri & \textbf{.8639}$\pm$.0024\textsuperscript{*} & \textbf{.9405}$\pm$.0009 & \textbf{.8711}$\pm$.0011 & \textbf{.7097}$\pm$.0004\textsuperscript{*} & \textbf{.7771}$\pm$.0024 & \textbf{.8632}$\pm$.0005\textsuperscript{*} & \textbf{.7410}$\pm$.0005\textsuperscript{*} & \textbf{.6843}$\pm$.0004\textsuperscript{*} & \textbf{.7850}$\pm$.0002\textsuperscript{*} & \textbf{.7437}$\pm$.0003\textsuperscript{*} \\
\bottomrule
\end{tabular}%
}
\vspace{-5px}
\end{table*}

In this section, we compare DuTri with multimodal and multi-scenario recommendation baselines. The experimental results are summarized in Table~\ref{tab:main_result}.

\noindent\textbf{Comparison with Multimodal Recommendation Baselines.} Results show that DuTri achieves higher mean AUC than the evaluated multimodal baselines in every scenario on both datasets. Although existing multimodal recommendation methods (e.g., Diff-MSIN and DMF) effectively model cross-modal interactions and can capture user-level modality preferences to some extent, they are primarily designed for single-scenario recommendation and lack explicit joint modeling of cross-scenario shared preferences and scenario-specific preferences. In contrast, DuTri learns collaborative preferences from user--scenario--item interactions and jointly uses user and scenario preferences to guide multimodal fusion. This allows the contributions of different modalities to adapt to the current user and scenario, incorporating preferences learned from cross-scenario interactions into multimodal modeling.

\noindent\textbf{Comparison with Multi-Scenario Recommendation Baselines.} The experimental results show that DuTri achieves higher mean WAUC than the evaluated multi-scenario baselines on both datasets, also outperforming Scenario MLP and AMS, which explicitly models modality scheduling. Existing multi-scenario recommendation methods (e.g., STAR and HMoE) effectively capture cross-scenario commonality and scenario differences through parameter sharing. However, under the early-fusion setting adopted in our evaluation, they incorporate multimodal information as additional inputs while retaining their original sharing architectures, without dedicated mechanisms for connecting collaborative preference learning with multimodal fusion. The competitive performance of Scenario MLP further indicates that sharing the prediction backbone after fusion does not necessarily outperform scenario-specific prediction networks under multimodal inputs. In contrast, DuTri shares user and item preference factors in the collaborative branch and uses scenario factors to modulate their interactions. It further uses user and scenario preferences to guide fusion in the multimodal branch and item preferences for cross-branch alignment, connecting cross-scenario collaborative modeling with multimodal content through preference representations.
\subsection{RQ2: Ablation Study}
\label{sec:experiments:rq2}
\begin{table*}[t]
\centering
\caption{Ablation results on both datasets, averaged over five random seeds. The highest mean in each column is boldfaced.}
\label{tab:ablation_result}
\setlength{\tabcolsep}{1pt}
\renewcommand{\arraystretch}{1.0}
\begin{tabular*}{\textwidth}{@{\extracolsep{\fill}}c*{10}{c}@{}}
\toprule
\multirow{3}{*}{Method} & \multicolumn{6}{c}{AntM$^2$C} & \multicolumn{4}{c}{Amazon} \\
\cmidrule(lr){2-7}\cmidrule(lr){8-11}
& \multicolumn{5}{c}{AUC$\uparrow$} & \multirow{2}{*}{WAUC$\uparrow$} & \multicolumn{3}{c}{AUC$\uparrow$} & \multirow{2}{*}{WAUC$\uparrow$} \\
\cmidrule(lr){2-6}\cmidrule(lr){8-10}
& S1 & S2 & S3 & S4 & S5 & & S1 & S2 & S3 & \\
\midrule
w/o Alignment & 0.8605 & 0.9365 & 0.8682 & 0.7075 & 0.7661 & 0.8597 & 0.7356 & 0.6831 & 0.7831 & 0.7398 \\
w/o CP Supervision & 0.8436 & 0.9356 & 0.8652 & 0.7059 & 0.7662 & 0.8568 & 0.7375 & 0.6758 & 0.7820 & 0.7397 \\
w/o RQ & 0.8626 & 0.9386 & 0.8692 & 0.7089 & 0.7768 & 0.8616 & 0.7399 & 0.6828 & 0.7833 & 0.7425 \\
w/o Gate & 0.8630 & 0.9401 & 0.8699 & 0.7089 & 0.7738 & 0.8621 & 0.7404 & 0.6807 & 0.7820 & 0.7421 \\
w/o UID Rep & 0.8553 & 0.9356 & 0.8695 & 0.7061 & 0.7753 & 0.8604 & 0.7393 & 0.6797 & 0.7803 & 0.7409 \\
w/o MM Rep & 0.8219 & 0.9358 & 0.8610 & 0.7056 & 0.7545 & 0.8528 & 0.7327 & 0.6724 & 0.7793 & 0.7356 \\
w/o CF Rep & 0.8626 & 0.9377 & 0.8700 & 0.7078 & 0.7756 & 0.8617 & 0.7373 & 0.6785 & 0.7794 & 0.7393 \\
w/o CF Br & 0.8319 & 0.9356 & 0.8632 & 0.7066 & 0.7665 & 0.8552 & 0.7337 & 0.6756 & 0.7813 & 0.7372 \\
w/o MM Br & 0.8011 & 0.9328 & 0.8619 & 0.7032 & 0.7555 & 0.8516 & 0.7300 & 0.6711 & 0.7763 & 0.7330 \\
MF Supervision & 0.8515 & 0.9354 & 0.8697 & 0.7070 & 0.7731 & 0.8603 & 0.7384 & 0.6718 & 0.7793 & 0.7390 \\
Pairwise Mod & 0.8615 & 0.9354 & 0.8707 & 0.7079 & 0.7731 & 0.8615 & 0.7394 & 0.6728 & 0.7803 & 0.7400 \\
USI Gate & 0.8605 & 0.9375 & 0.8682 & 0.7075 & 0.7761 & 0.8604 & 0.7399 & 0.6812 & 0.7800 & 0.7414 \\
\midrule
\textbf{DuTri} & \textbf{0.8639} & \textbf{0.9405} &  \textbf{0.8711} & \textbf{0.7097} & \textbf{0.7771} & \textbf{0.8632} & \textbf{0.7410} & \textbf{0.6843} & \textbf{0.7850} & \textbf{0.7437} \\
\bottomrule
\end{tabular*}
\vspace{-10px}
\end{table*}
We evaluate the contributions of DuTri's core components using the following variants: (1) ``w/o Alignment'' sets $\lambda_{\mathrm{cl}}=0$ to remove the alignment losses between item preferences and ID, textual, and visual representations; (2) ``w/o CP Supervision'' sets $\lambda_{\mathrm{CP}}=0$ to disable triadic CP Supervision while retaining all other components and objectives; (3) ``w/o RQ'' replaces quantized preferences with continuous preference factors; (4) ``w/o Gate'' replaces user--scenario-conditioned fusion with direct concatenation; (5) ``w/o UID Rep'', (6) ``w/o MM Rep'', and (7) ``w/o CF Rep'' remove the corresponding representations from the final prediction input while retaining all other components and auxiliary objectives; (8) ``w/o CF Br'' removes the entire collaborative branch and its associated auxiliary losses; (9) ``w/o MM Br'' removes the entire multimodal branch; (10) ``MF Supervision'' replaces triadic CP Supervision with user--item MF Supervision; (11) ``Pairwise Mod'' replaces user--scenario--item modulation with user--item modulation; and (12) ``USI Gate'' additionally conditions the fusion gate on item preferences. Table~\ref{tab:ablation_result} summarizes the results on AntM$^2$C and Amazon, where DuTri achieves the highest mean WAUC on both datasets.
The performance drops from removing alignment or CP Supervision support the contributions of cross-branch preference alignment and explicit triadic supervision, respectively. The lower performance of ``w/o RQ'' indicates the benefit of quantized preferences over continuous factors under the evaluated settings. The results of ``w/o Gate'' favor preference-conditioned weighted fusion over direct concatenation. Removing any prediction input reduces WAUC on both datasets, with ``w/o MM Rep'' exhibiting the largest drop, suggesting that all three representations provide additional predictive information and highlighting the contribution of fused item information. Removing either entire branch together with its auxiliary losses causes a larger WAUC drop than removing only its corresponding prediction input, supporting the contribution of each branch beyond its direct input to the prediction head. The lower performance of ``MF Supervision'' and ``Pairwise Mod'' supports incorporating scenario information into auxiliary preference supervision and collaborative representation modulation, respectively. Finally, ``USI Gate'' yields lower WAUC than DuTri on both datasets, supporting the choice not to condition the fusion gate on item preferences.
\subsection{RQ3: Alternative Modeling Strategies}\label{sec:experiments:rq3}
To examine alternative ways of incorporating multimodal information into multi-scenario recommendation, we construct four strategies using HierRec as a common backbone on AntM$^2$C: (1) ``Direct Input'' feeds fused item ID, textual, and visual representations into the backbone; (2) ``Quantized Input'' discretizes these representations and incorporates the corresponding codes as inputs, inspired by the semantic-ID strategy in RQ-GMM~\cite{tong2026rq}; (3) ``Shared-Specific Quant'' applies a shared quantization stage followed by a scenario-specific one, following the shared-bottom organization of shared and specialized components; and (4) ``Shared-Expert Quant'' combines shared codebooks with dynamically selected expert codebooks, inspired by the shared-expert tokenization in UniTok~\cite{hou2026tokenize}. These strategies adapt the above ideas to our setting. All strategies and DuTri use the same input features, data splits, and evaluation protocol.

As shown in Table~\ref{tab:modeling_strategies}, ``Quantized Input'' achieves lower mean WAUC than ``Direct Input'', suggesting that introducing quantized multimodal representations does not necessarily improve recommendation performance in this setting. Both ``Shared-Specific Quant'' and ``Shared-Expert Quant'' achieve higher mean WAUC than ``Direct Input'', supporting the consideration of cross-scenario sharing when organizing quantized representations. DuTri achieves the highest mean WAUC among the evaluated strategies. While the alternatives incorporate multimodal information through input representations and codebook-sharing mechanisms, DuTri learns a triadic preference structure from interaction feedback and uses it to connect collaborative and multimodal modeling through alignment and conditional fusion. Its overall advantage supports using cross-scenario collaborative preferences to guide multimodal modeling in the evaluated setting.
\begin{table}[t]
\centering
\caption{Multimodal multi-scenario modeling strategies on AntM$^2$C (five-seed averages). Comparison strategies use HierRec backbones. The highest mean in each column is boldfaced.}
\label{tab:modeling_strategies}
\setlength{\tabcolsep}{3pt}
\renewcommand{\arraystretch}{1.0}
\resizebox{\columnwidth}{!}{%
\begin{tabular}{@{}c*{6}{c}@{}}
\toprule
\multirow{3}{*}{Strategy}&\multicolumn{6}{c}{AntM$^2$C}\\
\cmidrule(lr){2-7}
&\multicolumn{5}{c}{AUC$\uparrow$}&\multirow{2}{*}{WAUC$\uparrow$}\\
\cmidrule(lr){2-6}
&S1&S2&S3&S4&S5&\\
\midrule
Direct Input&0.8182&0.9209&0.8680&0.7061&0.7701&0.8548\\
Quantized Input&0.8067&0.9145&0.8591&0.6911&0.7697&0.8461\\
Shared-Specific Quant&0.8093&0.9277&0.8679&0.7055&\textbf{0.7786}&0.8561\\
Shared-Expert Quant&0.8035&0.9283&0.8682&0.7052&0.7768&0.8560\\
\midrule
\textbf{DuTri}&\textbf{0.8639}&\textbf{0.9405}&\textbf{0.8711}&\textbf{0.7097}&0.7771&\textbf{0.8632}\\
\bottomrule
\end{tabular}%
}
\vspace{-10pt}
\end{table}
\subsection{RQ4: Hyperparameter Analysis}\label{sec:experiments:rq4}
In this subsection, we analyze the computational efficiency of DuTri and its sensitivity to key hyperparameters.

\begin{figure}[htbp]
    \centering
    \includegraphics[width=\linewidth]{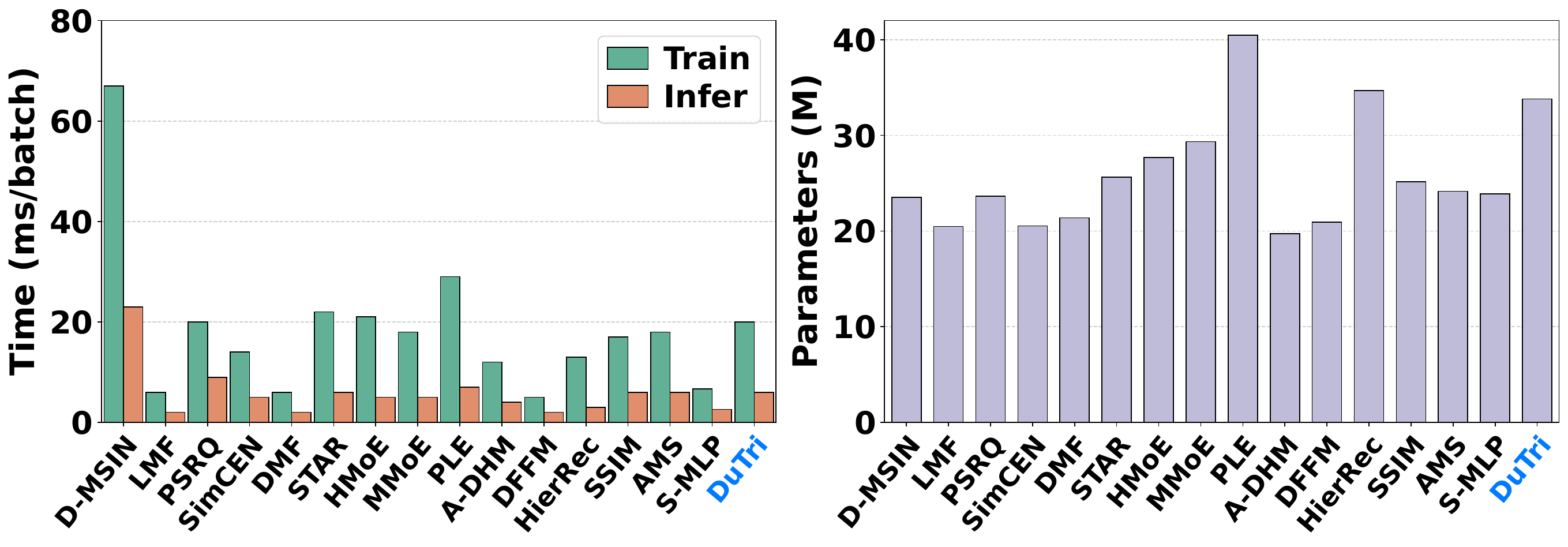}
    \Description[<Figure 5. Fully described in the text.>]{<A full description of Figure 5 can be found in Section 5.5.>}
    \caption{Efficiency comparison on AntM$^2$C.
    Left: per-batch training and inference times.
    Right: model parameter counts.}
    \label{fig:efficiency_antm2c}
    \vspace{-10px}
\end{figure}

\noindent\textbf{Efficiency Analysis.}
To evaluate the computational cost of DuTri, we compare per-batch training time, inference time, and parameter count on AntM$^2$C. As shown in Fig.~\ref{fig:efficiency_antm2c}, DuTri requires less training time than Diff-MSIN and PLE but more than AMS, while achieving inference time comparable to AMS. DuTri also has fewer parameters than PLE and HierRec but more than most other baselines, reflecting the additional parameter overhead introduced by preference factors and the associated representation learning modules. Together with its recommendation performance, these results indicate that DuTri achieves predictive gains at the cost of additional training time and model storage while maintaining inference efficiency comparable to AMS. Additional results on Amazon are provided in Appendix~\ref{app:efficiency_amazon}.
\begin{figure}[htbp]
\centering
\includegraphics[width=1.\linewidth]{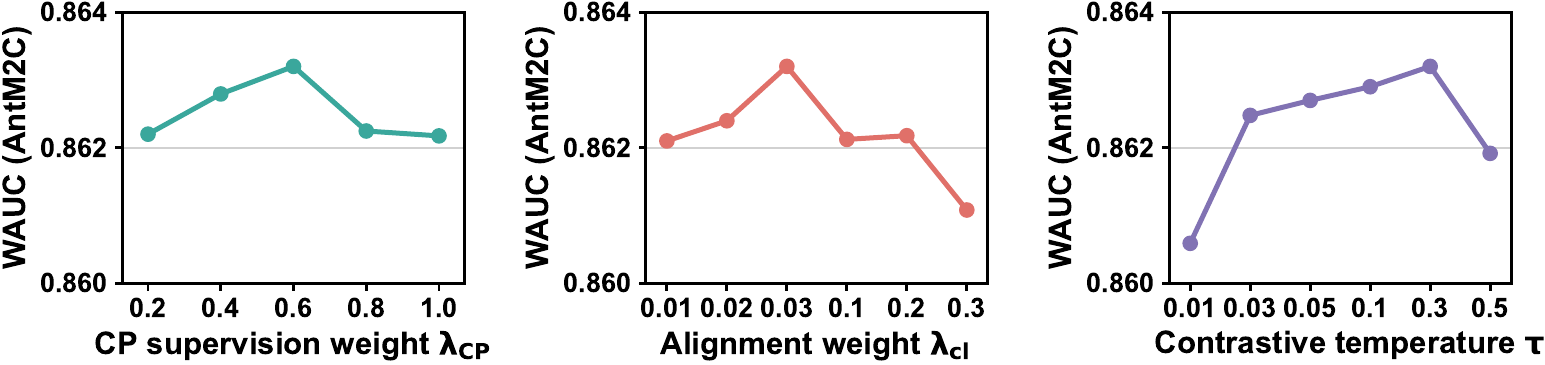}
 \vspace{-15pt}
 \Description[<Figure 4. Fully described in the text.>]{<A full description of Figure 4 can be found in Section 5.5.>}
\caption{Sensitivity to $\lambda_{\mathrm{CP}}$, $\lambda_{\mathrm{cl}}$, and $\tau$ on AntM$^2$C. Results are mean test WAUC over five random seeds.}
\label{fig:hyper}
\label{fig:hyper}
 \vspace{-10pt}
\end{figure}

\noindent\textbf{Hyperparameter Sensitivity Analysis.}
We examine the effects of the CP supervision weight $\lambda_{\mathrm{CP}}$, the alignment weight $\lambda_{\mathrm{cl}}$, and the contrastive temperature $\tau$ on AntM$^2$C. Hyperparameters and checkpoints are selected based on validation WAUC, and Fig.~\ref{fig:hyper} reports mean test WAUC. Performance remains relatively stable across the tested CP weights, whereas the alignment weight and temperature exhibit more pronounced performance peaks. Large alignment weights and excessively low or high temperatures are associated with lower performance. These results suggest that DuTri is relatively insensitive to the CP supervision weight within the tested range, while appropriate alignment weights and temperatures are needed for effective preference alignment.
\section{Conclusions}\label{sec:conclusions}
This paper presents DuTri, a dual-branch framework for multimodal multi-scenario recommendation that learns triadic collaborative preferences to guide multimodal fusion. Experiments on AntM$^2$C and Amazon demonstrate the best overall performance among the evaluated methods with competitive inference efficiency, while ablations and alternative strategy comparisons support its design.


\bibliographystyle{ACM-Reference-Format}
\bibliography{sample-base}

\appendix
\section{Additional Experimental Details}
\subsection{Motivational Comparison on Amazon}
\label{app:motivation_amazon}
\begin{figure}[htbp]
    \centering
    \includegraphics[width=1\linewidth]{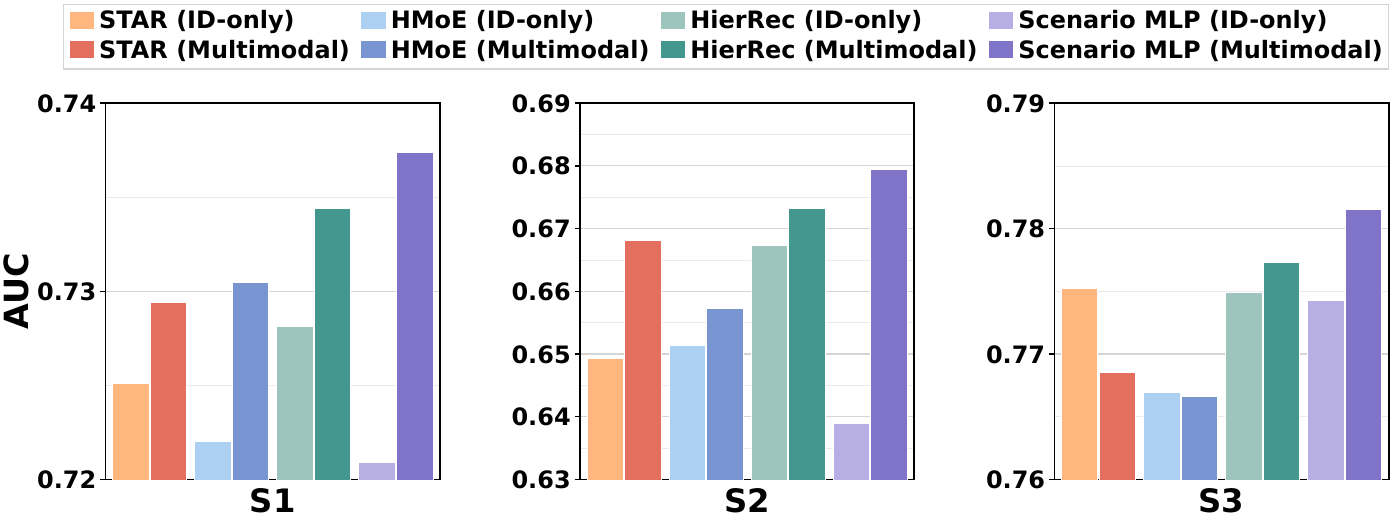}
    \Description[<Figure 6. Fully described in the text.>]{<A full description of Figure 6 can be found in Appendix A.1.>}
    \caption{Comparison of MSR backbones and Scenario MLP under ID-only and multimodal inputs on Amazon. Results report AUC for each scenario (S1--S3), averaged over five random seeds.}
    \label{fig:case_amazon}
    \vspace{-10pt}
\end{figure}

We repeat the comparison on Amazon using the same evaluation protocol, with the results shown in Fig.~\ref{fig:case_amazon}. As in the AntM2C comparison, we retain the shared ID embeddings and multimodal fusion layers, and replace the post-fusion MSR backbone with independent MLP prediction towers for individual scenarios.
\subsection{Additional Efficiency Analysis on Amazon}
\label{app:efficiency_amazon}
\begin{figure}[htbp]
    \centering
    \includegraphics[width=\linewidth]{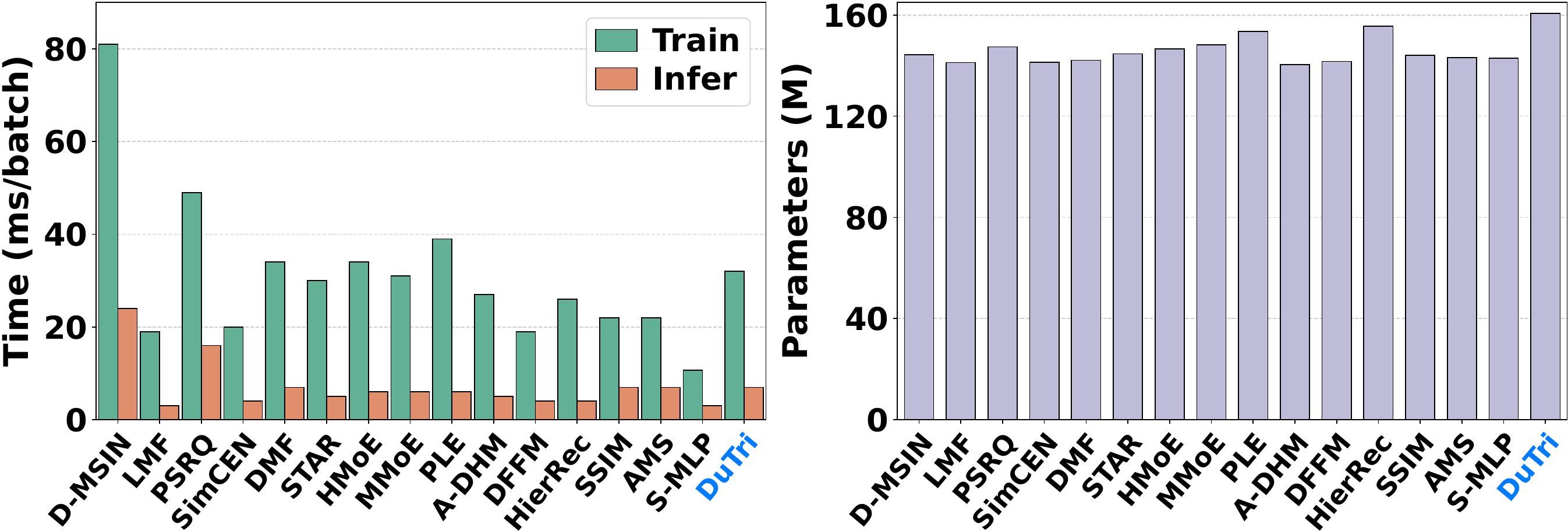}
    \Description[<Figure 7. Fully described in the text.>]{<A full description of Figure 7 can be found in Appendix A.2.>}
    \caption{Efficiency comparison on Amazon.
    Left: per-batch training and inference times.
    Right: model parameter counts.}
    \label{fig:efficiency_amazon}
\end{figure}

We compare per-batch training time, inference time, and parameter count on Amazon using the same evaluation protocol. As shown in Fig.~\ref{fig:efficiency_amazon}, DuTri requires less training time than Diff-MSIN, PSRQ, and PLE but more than AMS, and has the largest parameter count among the compared methods. Despite these additional training and parameter costs, its inference time remains comparable to AMS. Auxiliary losses and projection heads used exclusively for alignment are omitted during inference, so their training costs do not carry over to online prediction. Together with the recommendation results, these findings show that DuTri achieves higher overall WAUC with additional training and storage costs, reflecting a trade-off between predictive performance and computational efficiency.
\section{Experimental Settings}
\label{app:experimental_details}
\subsection{Dataset Statistics and Preprocessing}
\label{app:datasets}
Following the benchmark setup of AMS~\cite{hao2026ams}, we evaluate DuTri on AntM$^2$C\footnote{\url{https://www.atecup.cn/dataSetDetailOpen/1}} and Amazon\footnote{\url{http://jmcauley.ucsd.edu/data/amazon/links.html}}, which represent industrial business scenarios and category-based scenarios, respectively. The preprocessing procedures and feature configurations used in our experiments are described below. Table~\ref{tab:datasets} summarizes the processed dataset statistics.
\begin{table*}[t]
\centering
\caption{Statistics of the processed datasets.}
\begin{tabular}{c|c|c|c|c|c|c|c|c}
\specialrule{0.1em}{1pt}{1pt}
        \multicolumn{1}{c|}{\multirow{2}{*}{Dataset}} & \multicolumn{5}{c}{AntM$^2$C} & \multicolumn{3}{|c}{Amazon} \\
    \cline{2-9}
            \multicolumn{1}{c|}{} & S1 & S2 & S3 & S4 & S5 & S1 & S2 & S3 \\
    \hline
    Users     & 34,569   & 46,717   & 67,133   & 28,407   & 14,930   & 143,638  & 143,638  & 143,638   \\
    \hline
    Items     & 24,973   & 11,663   & 10,549   & 17,497   & 20,655   & 699,060  & 147,608  & 104,037   \\
    \hline
    Instances & 653,257  & 2,757,994 & 5,698,072 & 1,269,308 & 812,354 & 1,944,082 & 427,259 & 762,448   \\
    \hline
    Sparsity    & 99.92\%  & 99.49\%  & 99.20\%  & 99.74\%  & 99.74\%  & 99.99\%  & 99.99\%  & 99.99\%    \\
\specialrule{0.1em}{1pt}{1pt}
\end{tabular}
\label{tab:datasets}
\vspace{-10pt}
\end{table*}
\noindent\textbf{AntM$^2$C.} AntM$^2$C is an industrial multi-scenario CTR benchmark released by Alipay, containing click logs from five production scenarios, denoted by $\{s_k\}_{k=1}^{5}$. We use the official release and follow the multimodal feature extraction settings of prior work~\cite{huan2024exploring}. Low-frequency users have already been filtered from the released data, so no additional user filtering is applied. Textual and visual features are obtained using Multilingual BERT and Chinese-CLIP, with dimensions of 768 and 512, respectively. Missing images are represented by zero vectors. Following the temporal splitting protocol of AMS~\cite{hao2026ams}, we use the last four days for testing, the preceding two days for validation, and the remaining days for training. Specifically, interactions on or before August 3, 2023 are used for training, those from August 4--5 for validation, and those from August 6--9 for testing.

\noindent\textbf{Amazon.} Amazon contains user--item interactions, product metadata, and visual information. Following AMS~\cite{hao2026ams}, we use Sports and Outdoors, Movies and TV, and Books as scenarios $s_1$, $s_2$, and $s_3$, respectively. To establish user connections across scenarios, we retain users with interactions in all three categories and remove users with fewer than five interactions in total. Ratings above 3 are treated as positive feedback, and the remaining ratings as negative feedback. Within each scenario, each user's interactions are ordered chronologically. The most recent 10\% are used for testing, the preceding 10\% for validation, and the remaining 80\% for training. For textual features, we concatenate the product title, brand, category, and description, and encode the resulting text using \texttt{all-MiniLM-L6-v2}~\cite{reimers2019sentence}, yielding 384-dimensional representations. For visual features, we use the released 4,096-dimensional pre-extracted representations~\cite{ni2019justifying}. Missing visual representations are replaced by the mean of the available visual representations.
\subsection{Baseline Adaptation and Implementation}
\label{app:baselines}
We build on the baseline suite evaluated in AMS~\cite{hao2026ams} and additionally include DFFM, HierRec, and SSIM. AMS itself is also included as a baseline. The input adaptations and implementation settings are described below.

\noindent\textbf{Multimodal Methods.} Diff-MSIN, LMF, PSRQ, SimCEN, and DMF incorporate multimodal information but do not explicitly model scenario-specific interaction distributions. We train each method on the union of training interactions from all scenarios. Predictions are evaluated separately within each scenario, and the resulting AUCs are aggregated into WAUC. Scenario labels are used to group evaluation results.

\noindent\textbf{Multi-Scenario Methods.} STAR, HMoE, MMoE, PLE, AdaptDHM, DFFM, HierRec, and SSIM are adapted to multimodal inputs through early fusion. Following the concatenation-based adaptation used in AMS~\cite{hao2026ams}, item ID, textual, and visual representations are concatenated before being processed by each model's original multi-scenario architecture. This allows the existing parameter-sharing and interaction mechanisms to operate on multimodal inputs.

\noindent\textbf{AMS.} AMS models user- and scenario-dependent modality preferences through adaptive modality scheduling. The selected modalities are integrated by a multimodal fusion module and passed to a downstream multi-scenario backbone for prediction. We evaluate AMS using the same data splits and input features as DuTri.

\noindent\textbf{Scenario MLP.} For the motivational comparison, Scenario MLP retains the shared ID embeddings and multimodal fusion layers, while replacing the post-fusion multi-scenario backbone with independent MLP prediction towers for individual scenarios. The prediction towers share the same architecture but have separate parameters.

\noindent\textbf{Implementation Sources.} For the baselines included in both our study and AMS~\cite{hao2026ams}, we follow the same implementation settings. Specifically, STAR, MMoE, PLE, and AdaptDHM use the implementations provided by the Scenario-wise Rec benchmark, while Diff-MSIN, LMF, and SimCEN use their corresponding open-source repositories. HierRec, AMS, and SSIM are adapted from their publicly available implementations. For DMF, PSRQ, and HMoE, we follow the architectures and training procedures described in their original papers, consistent with the implementation approach adopted in AMS. DFFM is faithfully reimplemented according to the model architecture and training procedure specified in its original paper. All implementations are adapted to a common data input and evaluation pipeline. All baselines use the same textual and visual features as DuTri, and their hyperparameters are tuned independently on the validation set.

\subsection{Implementation Details and Hyperparameter Settings}
\label{app:implementation}

\noindent\textbf{Common Training Settings.} All models are trained using Adam with an ID embedding dimension of 128 and a batch size of 512. The learning rate is selected from $\{\texttt{1e-3},\texttt{1e-4},\texttt{1e-5}\}$, and the $\ell_2$ regularization coefficient is selected from $\{\texttt{1e-5},\texttt{1e-6},\texttt{1e-7}\}$. Hyperparameters are selected according to validation WAUC, where each scenario's AUC is weighted by its proportion of validation instances. The checkpoint with the highest validation WAUC is evaluated on the test set. Each experiment is repeated with five random seeds.

\noindent\textbf{DuTri Architecture.} The ID, textual, and visual 
representations are transformed into a common 64-dimensional space. Each scenario-specific prediction network has three hidden layers with dimensions $[1024,512,256]$. The prediction networks share the same architecture but have separate parameters. The CP rank is set to 16 on both datasets.

The user, scenario, and item factors are encoded using separate four-level residual quantizers with a quantization dimension of 32. Each level contains 256 codewords for user and item factors and 8 codewords for scenario factors. The commitment coefficient $\beta$ is set to 0.25. The representation alignment space has a dimensionality of 64.

\noindent\textbf{Auxiliary Objective Settings.} We select $\lambda_{\mathrm{CP}}$ from $\{0.2,\allowbreak 0.4,\allowbreak 0.6,\allowbreak 0.8,\allowbreak 1.0\}$. All three modalities share an alignment weight $\lambda_{\mathrm{cl}}$, selected from $\{0.01,\allowbreak 0.02,\allowbreak 0.03,\allowbreak 0.1,\allowbreak 0.2,\allowbreak 0.3\}$. The contrastive temperature $\tau$ is selected from $\{0.01,\allowbreak 0.03,\allowbreak 0.05,\allowbreak 0.1,\allowbreak 0.3,\allowbreak 0.5\}$.
\section{Collaborative Factor Quantization}
\label{app:quantization_details}
We quantize the user, scenario, and item CP factors with residual quantizers and train them jointly with the rest of DuTri. This section specifies the architecture, the forward computation, and the auxiliary objectives.
\subsection{Architecture}
For each entity type $A\in\{U,S,I\}$, the quantizer consists of a linear encoder that maps a CP factor $\mathbf v_A\in\mathbb{R}^{d_c}$ to $\mathbf z_A\in\mathbb{R}^{d_q}$, a residual quantizer with $L_A$ levels and per-level codebooks $\mathcal C_A^{l}$, and a linear decoder that maps the quantized representation back to $\mathbb{R}^{d_c}$. Each entity type has its own encoder, codebooks, and decoder. The CP rank is $d_c=16$ on both datasets, and the quantized dimension is $d_q=32$ for all entity types. Thus, each encoder maps 16-dimensional CP factors to a 32-dimensional latent space. The number of quantization levels is $L_A=4$ for all entity types. Each level of the user and item quantizers contains $K_A=256$ codewords, while each level of the scenario quantizer contains $K_A=8$ codewords. The commitment coefficient is $\beta=0.25$.
\subsection{Forward Computation}
Given $\mathbf v_A$, let $\boldsymbol\epsilon_A^{(0)}=\mathbf z_A$. For $l=1,\dots,L_A$, the quantizer selects a codeword and updates the residual:
\begin{equation}
\mathbf e_A^{(l)}=\arg\min_{\mathbf c\in\mathcal C_A^{l}}\big\|\boldsymbol\epsilon_A^{(l-1)}-\mathbf c\big\|_2^2,
\qquad
\boldsymbol\epsilon_A^{(l)}=\boldsymbol\epsilon_A^{(l-1)}-\operatorname{sg}\big(\mathbf e_A^{(l)}\big).
\end{equation}
The quantized representation is the sum of the selected codewords across levels:
\begin{equation}
\mathbf q_A=\sum_{l=1}^{L_A}\mathbf e_A^{(l)}.
\end{equation}
Straight-through estimation uses the quantized value in the forward pass and routes gradients to the encoder output:
\begin{equation}
\mathbf z_A^{\mathrm{ST}}=\mathbf z_A+\operatorname{sg}\big(\mathbf q_A-\mathbf z_A\big).
\end{equation}
The decoder maps $\mathbf z_A^{\mathrm{ST}}$ to the reconstructed factor $\widetilde{\mathbf v}_A^{\mathrm{ST}}$, whose forward value equals $\widetilde{\mathbf v}_A=\mathbf W_A^{\mathrm{dec}}\mathbf q_A+\mathbf b_A^{\mathrm{dec}}$. The reconstructed factor $\widetilde{\mathbf v}_A^{\mathrm{ST}}$ replaces $\mathbf v_A$ in all downstream computations, including the preference representations $\widetilde{\mathbf p}$, $\widetilde{\mathbf q}$, and $\widetilde{\mathbf r}$ in Section~\ref{sec:quantized_tri_modeling}.
\subsection{Training Objectives}
\noindent\textbf{Factor reconstruction.}
For a factor $\mathbf v$ with reconstruction $\widetilde{\mathbf v}$ and $\widetilde{\mathbf v}^{\mathrm{ST}}$, we define
\begin{equation}
\ell_{\mathrm{fac}}(\mathbf v)=\operatorname{MSE}\big(\widetilde{\mathbf v}^{\mathrm{ST}},\operatorname{sg}(\mathbf v)\big)+\operatorname{MSE}\big(\mathbf v,\operatorname{sg}(\widetilde{\mathbf v})\big),
\end{equation}
where the first term trains the encoder and decoder and the second term pulls the CP factor toward its reconstruction. We average over the entities of each type in the mini-batch and sum over types:
\begin{equation}
\mathcal L_{\mathrm{factor}}=\sum_{A\in\{U,S,I\}}\frac{1}{|\mathcal B_A|}\sum_{\mathbf v\in\mathcal B_A}\ell_{\mathrm{fac}}(\mathbf v),
\end{equation}
where $\mathcal B_U$ and $\mathcal B_I$ are the distinct users and items in the mini-batch, and $\mathcal B_S$ contains all scenarios. MSE averages over the dimensions.
\noindent\textbf{Quantization loss.} At level $l$ of entity type $A$, the codebook loss updates the selected codeword, while the commitment loss encourages the input residual to approach it:
\begin{equation}
\begin{aligned}
\ell_{\mathrm{code}}^{A,l} &= \operatorname{MSE}\big(\operatorname{sg}(\boldsymbol\epsilon_A^{(l-1)}), \mathbf e_A^{(l)}\big), \\
\ell_{\mathrm{cmt}}^{A,l} &= \operatorname{MSE}\big(\boldsymbol\epsilon_A^{(l-1)}, \operatorname{sg}(\mathbf e_A^{(l)})\big).
\end{aligned}
\end{equation}
Here, $\operatorname{MSE}$ averages over the factor vectors supplied to the quantizer and their latent dimensions. We average the losses over levels and sum them over entity types:
\begin{equation}
\mathcal L_{\mathrm{quant}} = \sum_{A\in\{U,S,I\}}\frac{1}{L_A}\sum_{l=1}^{L_A}\left(\ell_{\mathrm{code}}^{A,l}+\beta\,\ell_{\mathrm{cmt}}^{A,l}\right).
\end{equation}
The residual quantization objective combines factor reconstruction and quantization:
\begin{equation}
\mathcal L_{\mathrm{RQ}} = \mathcal L_{\mathrm{factor}} + \mathcal L_{\mathrm{quant}}.
\end{equation}
The resulting $\mathcal L_{\mathrm{RQ}}$ enters the overall training objective with unit weight.
\subsection{Batch Handling}
User and item factors are quantized once per distinct entity in the mini-batch and then gathered back to the sample level. Duplicated entities therefore share the same reconstruction and contribute once to $\mathcal L_{\mathrm{factor}}$ and $\mathcal L_{\mathrm{quant}}$. The scenario quantizer operates on the full set of scenario factors at every step. The CTR and CP losses are computed per sample.
\subsection{Codebook Usage and Discrete Codes}
Codeword usage counts are accumulated during training for monitoring only. We do not apply dead-code replacement. Training and inference both keep the full CP factor tables; the codebooks are not used to replace them with stored indices, and no index export or lookup is performed. The quantizer is therefore a discretized bottleneck in the forward pass, not a compressed representation.

\end{document}